\documentclass[onecolumn]{aa}
\usepackage{graphicx}
\usepackage{epsfig}
\usepackage{txfonts}
\begin{document}
   \title{Spectral Fits to the 1999 Aql X-1 Outburst Data}

   \author{T.J. Maccarone
          \inst{1}
          \and
          P.S. Coppi\inst{2}
          }

   \offprints{T. Maccarone}

   \institute{Astrophysics Sector, SISSA/ISAS, via Beirut n. 2-4,
              34014, Trieste, Italy\\ \email{maccarone@ap.sissa.it}
              \and Department of Astronomy, Yale University, P.O. Box
              208101, New Haven CT 06520-8101\\
              \email{coppi@astro.yale.edu} }

\date{} 

\abstract{We present analysis and spectral fits of the RXTE data from
the May/June 1999 outburst of Aql X-1.  These data include
observations in the rising portion of the hard state, in the soft
state, and in the falling portion of the hard state.  We show that the
data can be fit by a purely thermal Comptonization model for all the
observations, but that more complicated models cannot be ruled out.
Up to 60\% of the corona's power in the soft state may be injected
into non-thermal electrons.  The soft state data show approximately
constant optical depth and coronal temperatures over a range of
$\sim10$ in luminosity, while they show evidence for a reduction of
seed photon temperature with reduced luminosity and indicate that the
characteristic size of the seed photon emitting region is roughly
constant throughout the soft state.  The hard state before the soft
state shows a higher luminosity, higher optical depth, and lower
electron temperature than the hard state after the soft state.  We
find a reduction of the hard (30-60 keV) X-ray flux during a type I
burst and show that it requires a total corona energy reservoir of
less than $\sim 10^{38}$ ergs.  

 }
\authorrunning{Maccarone \& Coppi}

\maketitle

\section{Introduction} 

The lightcurves of X-ray transients have revealed an outburst cycle
consisting of several spectral states - an off/quiescent state where
the luminosity is low and the spectrum is difficult to measure, a
low/hard state, characterized by a power law with photon index
$\Gamma<2$, presumably coming from Comptonization of photons in a hot,
optically thin medium, a high/soft state, characterized by a
quasi-thermal spectrum, consistent with the emission from an accretion
disk with a peak temperature of $\sim 10^7$ K, often accompanied by a
weak power law tail, and a very high state, consisting of a strong
thermal accretion disk component plus a power law tail of comparable
or greater luminosity (see Nowak 1995 for a review of the
characteristics of the spectral states).  Additionally, the transition
between the hard and soft states is sometimes classified as a separate
``transition state''.

Two broad classes of mechanisms have been invoked to explain the
transitions between spectral states in black hole transients.  In one
mechanism, a two temperature accretion flow is proposed to explain the
hard state.  A transition radius between the cool, geometrically thin,
optically thick outer disk and the hot, optically thin, geometrically
thick inner corona changes with accretion rate, yielding a pure disk
spectrum at high luminosity and a Comptonized spectrum at lower
luminosity (see e.g Narayan \& Yi 1995; Meyer, Liu \&
Meyer-Hoffmeister 2000).

Alternatively, the differences between the spectral states may be
explained in terms of differing scale heights above the disk for
magnetically active hot regions in the hard state and the soft state
(see e.g. di Matteo, Celotti \& Fabian 1999; Nayakshin \& Svensson
2001).  The fraction of the luminosity supplied to the corona must
also change during the state transitions.  Relatively little work has
been done to explain why this should occur at different luminosities.

In neutron star systems, a propeller mechanism has sometimes been
invoked to explain state transitions (see e.g. Lamb, Pethick \& Pines
1973; Zhang, Yu \& Zhang 1998).  In this model, when the
magnetospheric radius of the neutron star (i.e. the radius where the
ram pressure of the accreting gas matches the magnetic pressure of the
accreting star) is outside the corotation radius of the neutron star,
then accreting matter is driven away from the neutron star by the
magnetic pressure and a hard state sets in.  The low luminosities
($\sim 10^{33}$ ergs/sec) seen in quiescence in Aql X-1 suggest that
some kind of outflow must be present in that state and that the
propeller is a likely mechanism (Campana et al. 1998).  What is not
certain is whether the magnetic field is large enough to cause the
onset of the propeller effect at the luminosities of the state
transitions.

Past studies of X-ray transients have focused on the broadband spectra
in quiescence (Yi et al. 1996), X-ray luminosities in quiescence (see
e.g. Garcia et al. 2001), and time delays between optical and X-ray
rises in outburst (see e.g. Jain et al. 2001).  Aql X-1 has been a
well studied source because it shows a fairly regular outburst cycle
with a periodicity of about 300 days.  It has been shown to be an
atoll source (Cui et al. 1998; Reig et al. 2000).  It has been the
subject of optical monitoring in quiescence (Jain 2001), and the
optical rise was used to trigger the pointed X-ray observations at a
lower flux than the sensitivity of the All Sky Monitor on the Rossi
X-ray Timing Explorer allows (Bradt, Rothschild, \& Swank 1993).  The
X-ray data triggered by the optical flare have allowed the analysis of
an X-ray lightcurve from a single outburst starting in the low/hard
state, entering the high/soft state, and returning to the low/hard
state; similar work to ours has been done recently for 4U 1704-44
without optical triggering (Barret \& Olive 2002).

\section{Observations} 

Aql X-1 was monitored by the RXTE pointed instruments during May and
June of 1999, caught early in the rising phase of its outburst due to
optical monitoring used to trigger the pointed observations (Jain
2001).  The observation times analyzed in this paper are listed in
Table 1.  A few additional observations have been omitted because
there were problems with the HEXTE background subtraction or because
the flux levels were too low for accurate spectral fits.  For each
observation, the data were extracted using the standard RXTE screening
criteria for earth elevation and time since the last South Atlantic
Anomaly passage.  The number of Proportional Counter Units turned on
varied, so the time interval used was chosen to maximize the total
integrated effective area times exposure time while still allowing for
a single set of PCUs to be used.  The HEXTE data are also extracted
according to the standard RXTE screening procedures.  For the
observations where Type I X-ray bursts are seen, the bursts are
excluded from the integration so as not to contaminate the spectrum.
The HEXTE data are rebinned to have three bins per channel from
channel 16 to channel 21, 5 bins per channel from channel 22 to 51, 15
bins per channel from channel 52 to 126, and 24 bins per channel from
channel 127 to 198.  The fits include PCA data from 3.0 to 20.0 keV
and HEXTE data from 17.0 to 190.0 keV, corresponding to the energy
ranges where the two instruments are generally considered reliable.

\begin{table*}
\begin{center}
{\bf Observations Analyzed for This Work}
\vskip 1cm

\begin{tabular}{cccccccccccc}
ObsId & StartDate & StartTime & StopDate & StopTime & JD-2451300\\
40049-01-02-00 & 15/05/99 & 17:23:04 & 15/05/99 & 18:37:15 & 13\\
40049-01-02-01 & 16/05/99 & 20:50:20 & 16/05/99 & 21:32:15 & 14\\
40049-01-02-02 & 17/05/99 & 22:29:22 & 17/05/99 & 23:06:15 & 15\\
40047-01-01-00 & 18/05/99 & 20:41:23 & 18/05/99 & 21:20:15 & 16\\
40047-01-01-01 & 19/05/99 & 22:24:23 & 19/05/99 & 23:04:15 & 17\\
40047-02-01-00 & 22/05/99 & 15:39:59 & 22/05/99 & 19:33:15 & 20\\
40047-02-02-00 & 23/05/99 & 18:59:08 & 23/05/99 & 22:59:15 & 21\\
40047-02-03-00 & 24/05/99 & 17:10:15 & 24/05/99 & 21:22:15 & 22\\
40047-02-03-01 & 26/05/99 & 17:11:36 & 26/05/99 & 17:34:47 & 24\\
40047-02-04-00 & 29/05/99 & 15:30:06 & 29/05/99 & 19:40:15 & 27\\
40047-02-05-00 & 31/05/99 & 15:29:23 & 31/05/99 & 19:58:15 & 29\\
40047-03-01-00 & 02/06/99 & 13:46:15 & 02/06/99 & 18:41:15 & 31\\
40047-03-02-00 & 03/06/99 & 15:25:26 & 03/06/99 & 20:49:15 & 32\\
40047-03-03-00 & 04/06/99 & 13:47:17 & 04/06/99 & 18:36:15 & 33\\
40047-03-04-00 & 05/06/99 & 07:21:37 & 05/06/99 & 09:57:15 & 34\\
40047-03-06-00 & 07/06/99 & 12:07:50 & 07/06/99 & 16:54:15 & 36\\
40047-03-07-00 & 08/06/99 & 12:06:42 & 08/06/99 & 13:51:05 & 37\\
40047-03-08-00 & 09/06/99 & 15:17:18 & 09/06/99 & 18:26:15 & 38\\
40047-03-09-00 & 11/06/99 & 15:15:53 & 11/06/99 & 16:32:15 & 40\\
\end{tabular}
\end{center}
\caption{The RXTE observations analyzed for this work.  Dates are
presented in DD/MM/YY format.}
\end{table*}

\section{Analysis}
\subsection{Data reduction}
The data are then fit with a thermal Comptonization model within XSPEC
10.0 (the eqpair model, with the electrons required to be put purely
into the thermal component - see section 3.2 below for a description
of the model), plus a Gaussian component to represent the
contributions of an iron line, multiplied by a photoelectric
absorption component (wabs).  The purely thermal version of this model
is chosen as the simplest one that fits the data well and gives some
insight about which physical parameters are varying.  A hybrid
thermal/non-thermal version of the model is also fit to the data later
on.  The neutral hydrogen column density is frozen to
$3.4\times10^{21}$ cm$^{-2}$, the Galactic value along the line of
sight to Aql X-1.  A 1\% systematic error is added to all channels
from both the PCA and the HEXTE.  The overall normalization of the
HEXTE data is allowed to float and varies by about 30\%.  The Gaussian
component is required to have an energy between 6.2 and 7.0 keV and a
physical width of less than 1.2 keV.  For most of the observations,
the presence of the line is only marginally statistically significant.
Given the slightly larger systematic errors near the iron line than in
the rest of the RXTE spectrum and that these errors tend to make the
measured flux in this region larger than that measured with other
instruments, we view the measurements of the line with skepticism,
omit the parameters of the line measurements from the tables and
refrain from discussing any implications they might have; while the
formal statistical significance of the line is high, it is possible
that the fits are responding to errors in the response matrix.

\subsection{The Comptonization model}
 The EQPAIR model calculates a self-consistent temperature and pair
optical depth for the hot electron corona.  It allows for injection of
energy into the coronal electrons in both a Maxwellian distribution
(``thermal Comptonization'') and a power law distribution
(``non-thermal Comptonization'').  It then solves for a final electron
distribution based on the relative heating and cooling rates of the
electrons due to radiative processes and Coulomb interactions, so the
electron temperature is not an explicit parameter, but is instead
computed self-consistently.  It allows for Compton reflection, but
does not self-consistently compute the line emission from reflection.
It assumes a spherical geometry with the seed photons initially
distributed uniformly.  We choose this model because it is versatile
(allowing for both thermal and non-thermal injection and allowing for
Compton reflection) and runs relatively quickly given its complexity.

The basic parameters of the thermal Comptonization component of the
model are a thermal electron compactness (equal to the luminosity
injected into the Maxwellian component of the hot electron
distribution times the Thomson cross-section and divided by the
product of the electron rest energy, the speed of light and the radius
of the corona), and optical depth, and a seed photon temperature.  The
temperature of the corona is an implicit one determined by the energy
balance between the heating rate and cooling rate of the corona.  The
compactness of the seed photon distribution is frozen to unity, so the
thermal electron compactness, $\ell_{th}$ is really the ratio of the
coronal luminosity to the seed photon luminosity.  Barring an
independent estimate of the corona's size, the strongest constraints
on the total compactness will come from annihilation line
measurements, which are not possible given that our observations
extend only to 190 keV.  For a more detailed description of the model,
we refer the reader to Coppi 1999.

\subsection{Brief overview of results}
The results of the spectral fitting are presented in Table 2 and
plotted in Figure 1.  The important implicit parameters - the
luminosity in the soft component, the luminosity in the corona, and
the temperature of the corona - are plotted in Figure 2.  The fluxes
are converted to luminosities assuming a distance to Aql X-1 of 2.5
kpc (Chevalier et al. 1999), although we note that this distance
measurement has been disputed by Rutledge et al. (2001), who claim a
minimum distance of 4 kpc and a most likely distance of 4.7 kpc.  The
relative luminosities of the soft component and coronal component are
computed by multiplying the total luminosity by the fraction of the
total compactness in each component.  From inspection of Table 2, it
seems that the source entered the soft state on May 22 and remained
there through June 7 (JD 2451320 through 2451338).  These dates could
be in error by a few days due to data gaps.  In the hard state, the
optical depths range from $\sim$ 1-4 and the coronal temperatures
range from $\sim$ 10-200 keV (a typical hard state spectrum is plotted
in Figure 2).  In the soft state, the coronal optical depths are
$\sim$ 10 and the temperatures are $\sim$ 2.5 keV.  Two different
model fits to a typical soft state spectrum are presented in Figure 3,
while a typical hard state spectrum is plotted in Figure 4.

\begin{table*}
\begin{center}
{\bf Best Fit Parameters for the Thermal Model}
\vskip 1cm
\small
\begin{tabular}{cccccccccccc}
ObsID & $kT_{bb}$ & $\ell_{th}$ & $\tau$ & $L$ & $\chi^2/\nu$\\
40049-01-02-00 & $190\pm^{80}_{30}$ & $3.69\pm^{.08}_{.06}$ & $2.40\pm^{.07}_{.05}$ & 2.3$\times10^{36}$ & 1.73\\
40049-01-02-01 & $260\pm^{18}_{12}$ & $3.34\pm^{.07}_{.06}$ & $2.39\pm^{.13}_{.06}$ & 3.1$\times10^{36}$ & 1.75\\
40049-01-02-02 & $260\pm^{18}_{12}$ & $3.34\pm^{.07}_{.06}$ & $2.39\pm^{.13}_{.06}$ & 3.9$\times10^{36}$ & 1.36\\
40047-01-01-00 & $760\pm^{70}_{130}$ & $1.60\pm^{.25}_{.07}$ & $3.02\pm^{.62}_{.40}$ & 4.2$\times10^{36}$ & 0.72\\
40047-01-01-01 & $700\pm^{60}_{100}$ & $1.41\pm^{.06}_{.08}$ & $4.23\pm^{.23}_{.18}$ & 4.2$\times10^{36}$ & 0.84\\
40047-02-01-00 & $620\pm^{50}_{100}$ & $0.55\pm^{.08}_{.08}$ & $10.12\pm^{.06}_{.08}$ & 5.5$\times10^{36}$ & 1.00\\
40047-02-02-00 & $560\pm^{80}_{40}$ & $0.71\pm^{.01}_{.01}$ & $11.45\pm^{.19}_{.20}$ & 8.1$\times10^{36}$ & 0.63\\
40047-02-03-00 & $580\pm^{140}_{70}$ & $0.55\pm^{.13}_{.07}$ & $8.82\pm^{.68}_{.60}$ & 6.1$\times10^{36}$ & 0.84\\
40047-02-03-01 & $590\pm^{90}_{80}$ & $0.62\pm^{.10}_{09}$ & $10.70\pm^{.6}_{.6}$ & 7.8$\times10^{36}$ & 0.55\\
40047-02-04-00 & $510\pm^{110}_{**}$ & $0.79\pm^{.16}_{**}$ & $12.10\pm^{0.5}_{0.5}$ & 6.1$\times10^{36}$ & 0.61\\
40047-02-05-00 & $172\pm^{15}_{12}$ & $1.06\pm^{.02}_{.02}$ & $9.88\pm^{.12}_{.24}$ & 2.3$\times10^{36}$ & 0.99\\
40047-03-01-00 & $340\pm^{120}_{20}$ & $0.85\pm^{.02}_{.09}$ & $9.95\pm^{.25}_{.15}$ & 3.1$\times10^{36}$ & 0.84\\
40047-03-02-00 & $560\pm^{10}_{10}$ & $0.55\pm^{.01}_{.01}$ & $8.38\pm^{.10}_{.12}$ & 2.1$\times10^{36}$ & 1.68\\
40047-03-03-00 & $140\pm^{300}_{**}$ & $1.23\pm^{.5}_{.**}$ & $10.31\pm^{.34}_{.26}$ & 1.9$\times10^{36}$ & 1.10\\
40047-03-04-00 & $420\pm^{**}_{60}$ & $0.83\pm^{.03}_{.03}$ & $12.29\pm^{.40}_{.26}$ & 1.7$\times10^{36}$ & 0.80\\
40047-03-06-00 & $350\pm^{210}_{120}$ & $0.93\pm^{.03}_{.06}$ & $11.86\pm^{.50}_{.43}$ & 1.1$\times10^{36}$ & 1.06\\
40047-03-07-00 & $140\pm^{**}_{**}$ & $1.29\pm^{.28}_{.07}$ & $10.99\pm^{.57}_{.44}$ & 7.5$\times10^{35}$ & 0.67\\
40047-03-08-00 & $740\pm^{40}_{35}$ & $2.02\pm^{.16}_{.14}$ & $0.84\pm^{.10}_{.07}$ & 6.1$\times10^{35}$ & 0.58\\
40047-03-09-00 & $660\pm^{40}_{35}$ & $2.52\pm^{.38}_{.22}$ & $0.69\pm^{.14}_{.11}$ & 3.3$\times10^{35}$ & 0.80\\

\end{tabular}
\end{center}
\caption{The best fit parameter values for the thermal Comptonization
fits to the Aql X-1 outburst data.  Double asterisks represent
parameters whose error bars were either too large to be fit (a factor
of three or more from the best fit value), or were large enough to
exceed the hard bounds placed on the parameters.  There are 59 degrees
of freedom for each observation.  The columns are: ObsID - the RXTE
observation ID number, $kT_{bb}$ - the seed photon temperature in eV,
$\ell_{th}$ - the thermal compactness of the corona, $\tau$ - the
optical depth of the corona, $L$ - the total luminosity of the source,
assuming a 2.5 kpc distance, and $\chi^2/\nu$, the reduced $\chi^2$
from the spectral fits.}
\end{table*}

\section{Discussion}
\subsection{Robustness of fits and alternative models}

Previous work has attempted to fit the data as coming from a disk
blackbody model plus a Comptonized blackbody (Mitsuda et al. 1984;
Barret et al. 2000).  We find that for some hard state observations,
the addition of a second blackbody component can give a slight
improvement to the goodness of fit, and results in a slightly higher
coronal temperature and lower optical depth, as well as significant
changes in the iron line flux.  However, since the fits are
satisfactory with a single Comptonized blackbody, and the additional
blackbody generally represents less than $\sim$ 20\% of the 2-200 keV
flux, we use the simpler single Comptonized blackbody model.  In
general, RXTE is not capable of distinguishing between the different
models for the soft flux or between the disk and boundary layer
components because it has relatively poor spectral resolution at
energies below 5 keV and has no coverage below 2 keV.  Observations
from instruments such as BeppoSax or XMM-Newton are more useful for
making such measurements.  Furthermore, we note that the soft photon
temperatures derived with RXTE may be subject to significant
systematic errors because they are determined predominantly by the
lowest energy channels which have relatively high systematic errors.

Under the assumption that the EQPAIR model is a correct
description of the data, it is potentially sensitive to changes in the
seed photon temperature even when the seed photon distribution has
dropped off significantly by the lower bound of the observed energy
range.  The reason is that the broad bump of the seed photons must
join smoothly to the "power law" that comes out of analytic
calculations assuming a delta function seed photon distribution.  Thus
there is some curvature at low energies which deviates from the
idealized power law description, and this curvature constrains the
seed photon temperature.  In fact, this is a generic property of
numerically computed Comptonization models.  Similar results have been
found by Nowak, Wilms and Dove (2002) for GX 339-4 fits.  We caution
that systematic errors in the RXTE response matrix and in the neutral
hydrogen column are likely to cause systematic errors in the
temperature measurements for all sources.  For neutron star sources,
the possible existence of two seed photon components (one from the
star's surface and one from the disk) complicates matters further.  We
do believe that systematic trends in the seed temperatures are
meaningful (i.e. where there is statistically significant evidence
that the temperature in observation A is higher than that in
observation B, we do believe that it is, in fact, higher), but we do
not believe that RXTE is the best mission for making actual
measurements of the temperature.

We also attempt to fit the data with thermal Comptonization plus
Compton reflection.  We find that the fits are unconstrained when the
channels between 6 and 10 keV are ignored, implying that the fit is
sensitive only to the edges and not to the curvature of the continuum.
Since the edges depend sensitively on the RXTE response matrix and on
whether one or two seed blackbody photon distributions are included,
we believe little can be said about whether Compton reflection is
important in these data.

The spectra in the soft state fit the canonical high optical depth,
low temperature corona for a Comptonized blackbody that has been seen
in other neutron star sources (e.g. Di Salvo et al. 2001).  It is of
some interest to note that this spectral form matches both the Z
source GX 349+2 and the atoll source Aql X-1 in its soft state,
providing further evidence that the atoll/Z-source designation is
somewhat arbitrary (see Muno, Remillard \& Chakrabarty 2002).  Within
the soft state, the optical depth, coronal temperature, and seed
photon temperature do not to show any trends, except that the lowest
seed photon temperatures seem to be found after the large luminosity
drop on JD 2451329.  The best fitting disk temperature seems to drop
by a factor of about 2 as the source drops in luminosity by a factor
of about 10. Specifically, the first five observations in the soft
state have disk temperatures of about 550 eV and luminosities in the
range of $6-8\times10^{36}$ ergs/sec, while the later observations in
the soft state generally have temperatures around 300 eV, with
luminosities in the range $1-3\times10^{36}$ ergs/sec.  This presents
evidence that the size of the region from which the seed photons come
does not vary by more than a factor of about 2 in the soft state, as
for a constant sized emission region, $T_{bb}\propto L^{1/4}$.
Previous work on the black hole systems GRO J 1655-40 (Sobczak et
al. 1999) and XTE J 1550-564 (Sobczak et al. 2000) has shown that
their inner disk radii are likely to remain constant through the
high/soft state as well.  We stress that while the actual temperatures
(and hence the actual size of the emission region) are susceptible to
rather large systematic errors, the trends in the temperature should
be much more reliable.

We also fit the soft state data with a hybrid thermal/non-thermal
Comptonization model to determine whether there might be a power law
tail to the electron energy injection function.  We find that the data
for a typical soft state observation (observation 40047-03-01-00) are
consistent with up to 60\% of the coronal luminosity injected into
non-thermal electrons (i.e. electrons with a power law distribution),
but the data are not good enough to distinguish between the models.
The plots of the pure thermal electron distribution fit and the hybrid
distribution fit are shown in Figure 3, which shows weak evidence for
an extended tail above 20 keV.  We note that it does not appear that
either model fits the HEXTE data well, but that the errors on the
HEXTE data are quite large and that given the location of Aql X-1 in
the disk of the Galaxy, activity of a foreground/background source
cannot be ruled out as a possible cause for this extra hard X-ray
emission.  To date, such extended power law tails have been seen in
black hole sources (see e.g. Gierlinski et al. 1999) and in Z source
neutron stars (D'Amico et al. 2001; di Salvo et al. 2001), but not in
atoll sources.  These tails have been proposed as coming from
non-thermal electron distributions (see e.g. Coppi 1999) or bulk
motion Comptonization from the inner accretion flow (see
e.g. Papathanassiou \& Psaltis 2001).  It has been argued that the
presence of an extended tail is a definitive signature of strong
gravity and hence of the presence of a black hole (Titarchuk \&
Zannias 1998).  Since it could be argued that the non-thermal tails
seen in Z sources are related to the stars' relatively strong magnetic
fields (see e.g. Miller, Lamb \& Psaltis 1998), observations or strong
upper limits of non-thermal tails in atoll sources are needed to
determine which, if any, of the theoretical pictures is correct.  The
present data are not sufficient for any strong constraints.  A long
integration with a sensitive hard X-ray/soft $\gamma-$ray telescope
such as INTEGRAL should be able to resolve the question of the
nonthermal tail.

\begin{figure*}
\centerline{\epsfig{figure=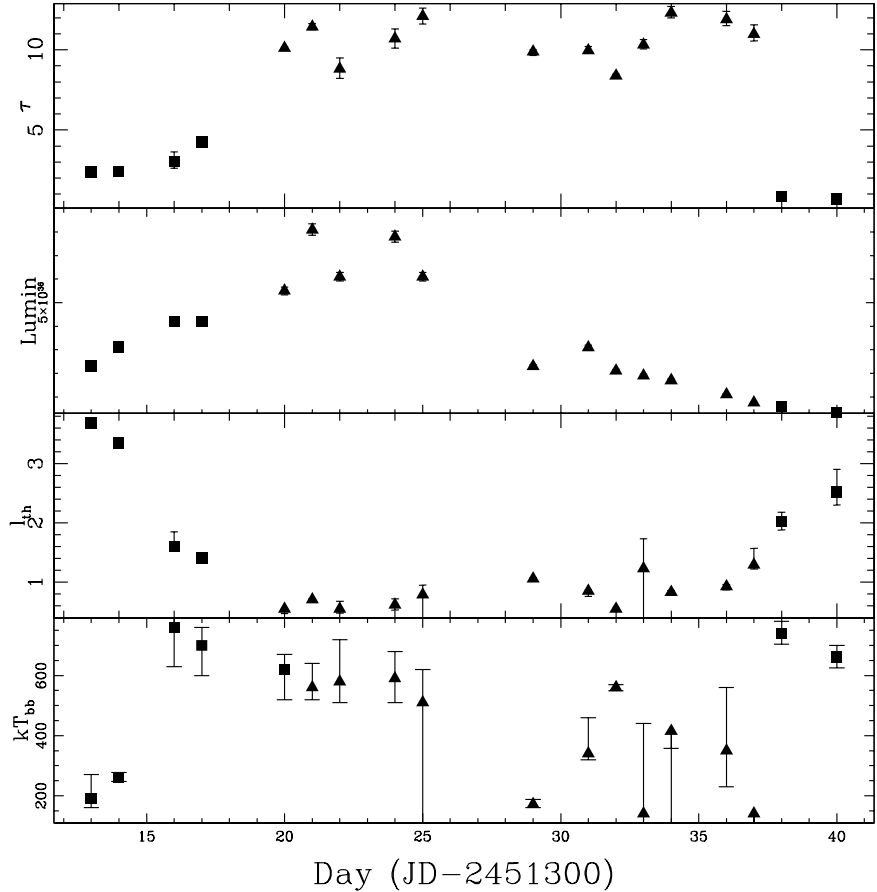,width=13 cm}}
\caption{The spectral fit parameters plotted versus the observation
date.  From top to bottom: the optical depth, the luminosity in
ergs/sec/cm$^2$, the thermal compactness, and the blackbody temperature
in eV.  In all panels, the filled squares show the observations we have defined to the hard state and the filled triangles show the observations we have defined to be the soft state.}
\end{figure*}

\begin{figure*}
\centerline{\epsfig{figure=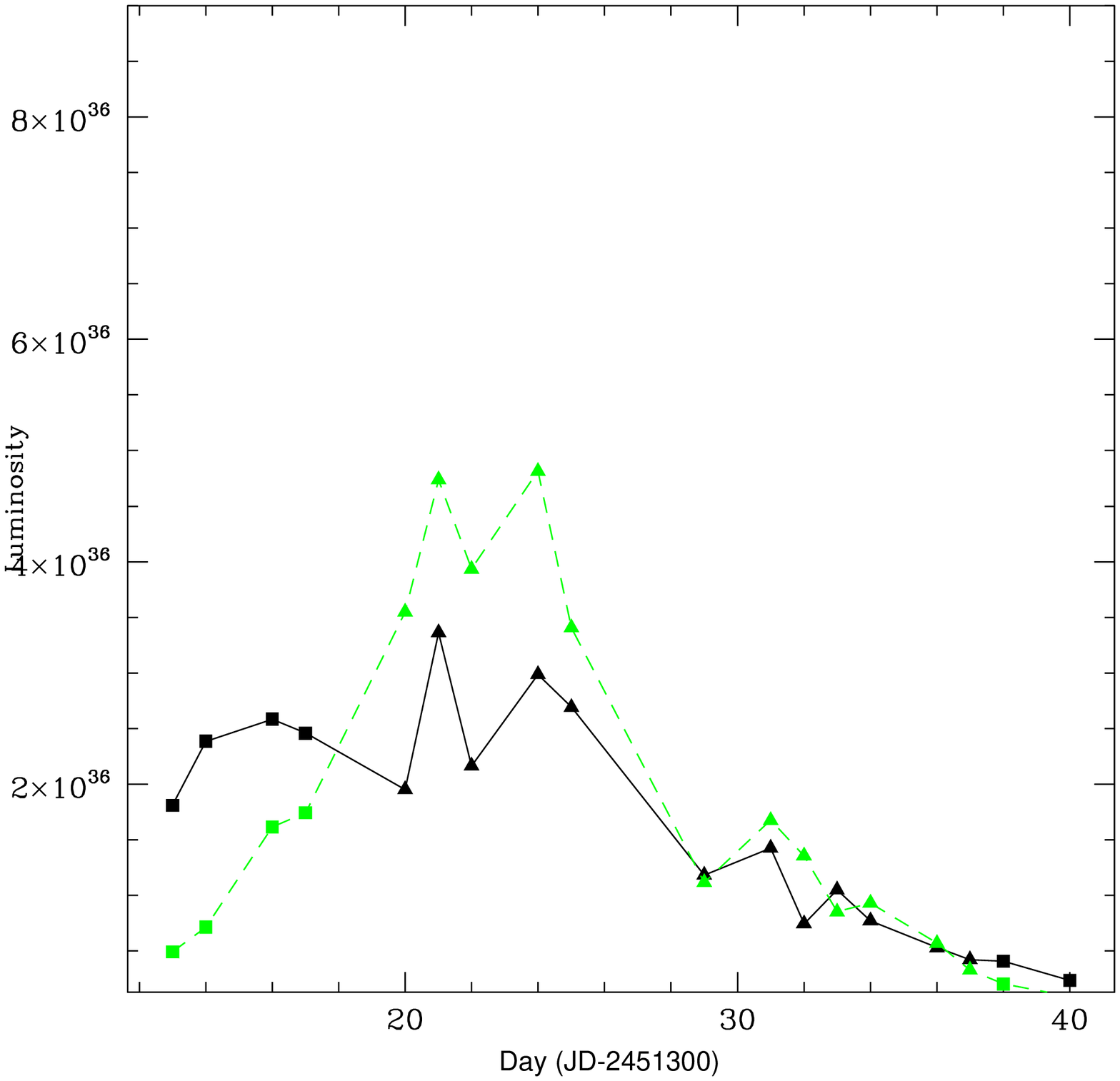,width=6.2 cm,angle=0}\epsfig{figure=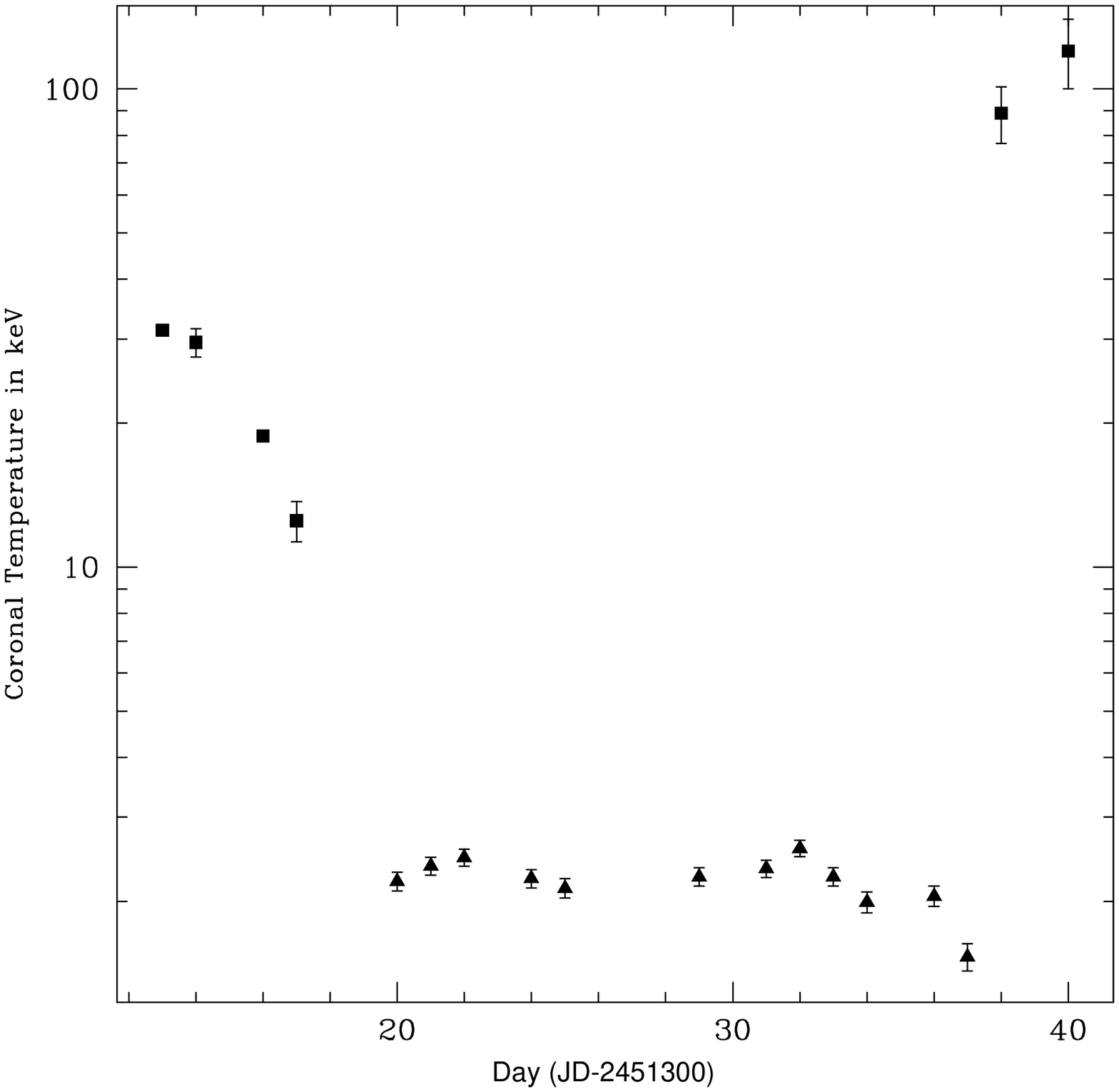,width=6.2 cm,angle=0}}
\caption{(a)The upscattered luminosity (solid line) and the seed
photon luminosity (dashed line) plotted versus day. (b) The coronal
temperature plotted versus day.In both plots, the filled squares show
the observations we have defined to the hard state and the filled
triangles show the observations we have defined to be the soft state.}
\end{figure*}              

\begin{figure*}
\centerline{\epsfig{figure=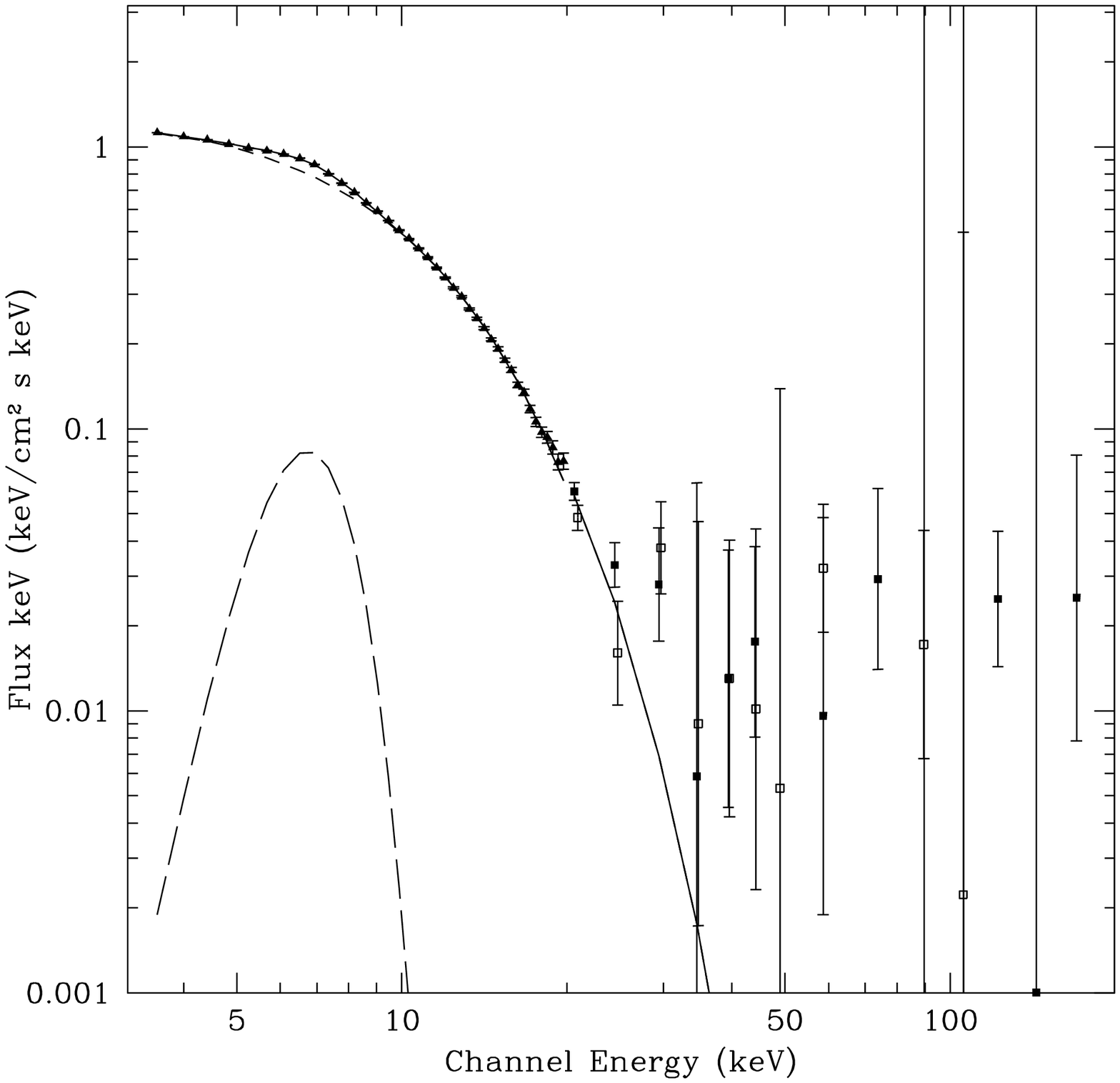,width=6.2 cm,angle=0}\epsfig{figure=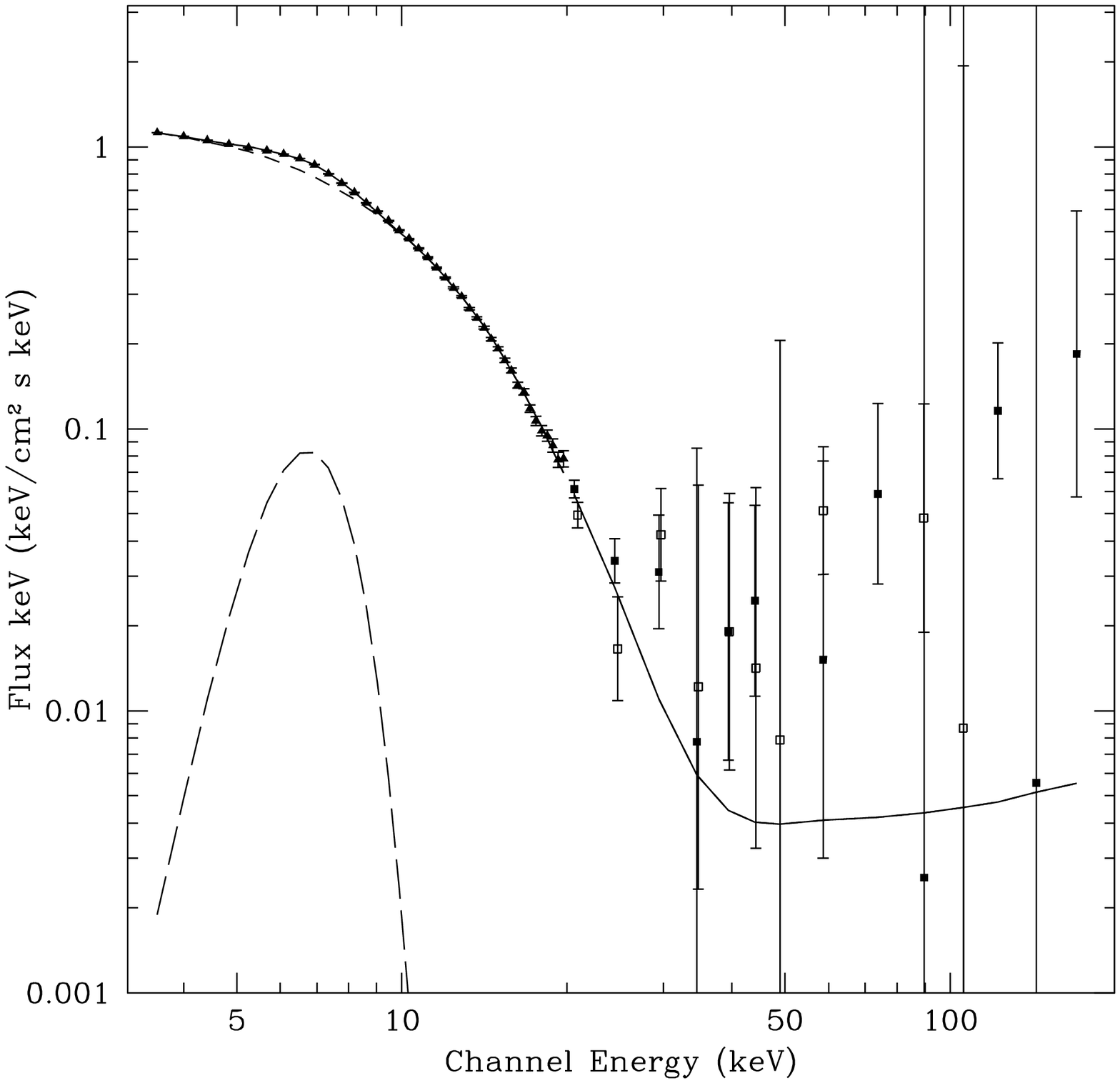,width=6.2 cm,angle=0}}
\caption{The fits to soft state data (ObsID 40047-03-01-00) for Aql
X-1.  The figure to the left is that for the thermal model and to the
right is for the hybrid model.  The triangles are the PCA data,
multiplied by 0.7 to match with the HEXTE data.  The filled squares
are HEXTE cluster 0 and the open squares are HEXTE cluster 1.  The
solid line represents the sum of the model components, the
short-dashed line represents the EQPAIR contribution and the
long-dashed line represents the Gaussian ``iron line.''  Where no
error bars are visible, the errors are smaller than the size of the
data point.}
\end{figure*}

\begin{figure*}
\centerline{\epsfig{figure=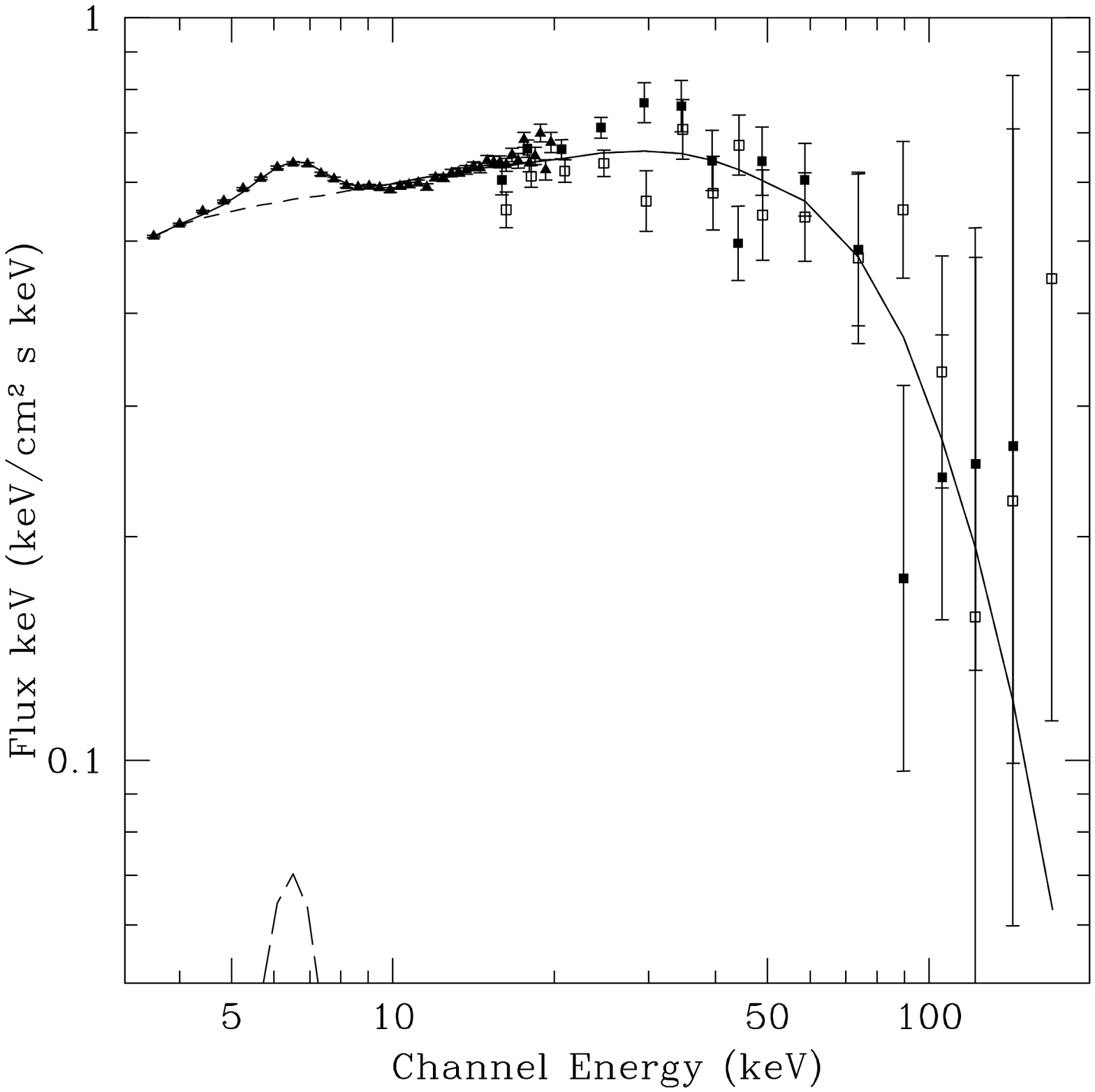,width=13 cm,angle=0}}
\caption{The fit to the hard state data for Aql X-1.  The observation
chosen is 40049-01-02-02, shortly before the transition to the soft
state.  The triangles are the PCA data, multiplied by 0.7 to match
with the HEXTE data.  The filled sqaures are HEXTE cluster 0 and the
open squares are HEXTE cluster 1.  The solid line represents the sum
of the model components, the short-dashed line represents the EQPAIR
contribution and the long-dashed line represents the Gaussian ``iron
line.'' Where no error bars are visible, the errors are smaller than
the size of the data point.}
\end{figure*}

We find results for the hard state observations that are consistent
with other work (e.g. Barret et al. 1999).  However, we note some
systematic differences between the rising hard state and the falling
hard state.  Specifically, on the way into the soft state, the corona
has a higher optical depth, higher compactness ratio, and a lower
temperature than on the way out of the soft state.  The luminosity of
the source is substantially higher in rising portion of the hard state
than in the falling portion of the hard state.  The blackbody
temperatures in the rising portion of the hard state are lower than in
the falling portion.  The spectral fit to a typical low/hard
state observation is shown in Figure 4.

The differences in coronal parameters of the rising and falling hard
states are likely to be straightforward consequences of the luminosity
difference - at higher luminosities the corona will be cooler and have
a larger optical depth.  Such results have been predicted
theoretically (see e.g. Esin, McClintock \& Narayan 1997). They are
furthermore suggested in the black hole system XTE J 1550-564 by the
hardening of the 20-100 keV spectrum as the source drops in luminosity
through its low/hard state (Sobczak et al. 2000).That the luminosity
is higher on the way into the soft state than on the way out indicates
a hysteresis effect in the state transitions.  This observation and
similar observations for several black hole soft X-ray transients are
discussed in another paper (Maccarone \& Coppi 2002).  We briefly note
that this observation argues strongly against the propeller effect as
the sole cause of the state transitions in Aql X-1, as propeller
models predict a constant state transition luminosity, and that a
single model for black hole and neutron star state transitions is
instead suggested.

\section{Observations of a Type I burst During the Hard State}
We have also observed a type I X-ray burst during Observation
40047-01-01-01, shortly before the transition from the hard state to
the soft state.  During this burst, the PCA count rate rises by a
factor of about 10, the HEXTE count rate from 15-30 keV rises by a
factor of about 20\% and the HEXTE count rate from 30-60 keV drops by
a factor of about 2.  The drop in the 30-60 keV band is significant at
only the 2 $\sigma$ level, but its occurrence simultaneously with the
burst eliminates the possibility of ``hidden trials'' and allows the
one-sided confidence interval of the result, i.e. 98\%, to be taken
literally.  The light curve of this burst in the two HEXTE energy
bands, binned on a 64 second timescale, is shown in Figure 5.  The
burst began at t=169771900 seconds in RXTE mission units, peaked 9
seconds later, and lasted about 100 seconds before the flux returned
to the pre-burst level.

The observation of a type I X-ray burst during the hard state provides
a probe for the energy content of the corona.  The X-ray burst
increases the soft photon luminosity by a factor of $\sim10$,
presumably without increasing the power supplied to the electrons in
the corona.  A uniform tempertature and density spherical corona will
have a total energy content of $\frac{4\pi}{3}R_{cor}^3 n_e k_BT$.
Thus for a uniform temperature and density corona with $\tau>1$,
$E_{cor}$, the energy content of the corona is:
\begin{equation}
E_{cor}= 5\times 10^{31}(y/2)(\tau/5)^{-1}(\frac{R_{cor}}{100
\rm{km}})^2 \rm{ergs},
\end{equation}
where $y$ is the Compton $y$ parameter ($y=4kT
\rm{max}(\tau,\tau^2)/m_ec^2$), $\tau$ is the optical depth of the
corona, and $R_{cor}$ is the radius of the corona, and where the
leading coefficient is the numerical evaluation of
$\frac{2\pi}{15}\frac{(100 km)^2}{\sigma_T} m_e c^2$ in units of ergs.
The additional seed photons in the burst should provide for additional
cooling of the corona.  If this additional cooling takes away an
amount of energy that is comparable to the total energy content of the
corona, then the temperature of the corona should drop significantly.
Such a temperature drop is seen, in the form of the factor of 2
decrease in the 30-60 keV flux during the burst.  Since the fluence of
the burst is large ($\sim 3 \times 10^{38}$ ergs), we can place only
the weak limit on the corona size that it must be less than $\sim$ 1
light-second in size if it has uniform temperature and density.  The
sensitivity of HEXTE is not sufficient to obtain strong upper limits
on the corona's size from a single burst.  Stacking bursts so that a
statistically significant change in the HEXTE count rate can be
detected more quickly (and hence after less total soft fluence) should
be a topic for future work.   If future observations measure an
energy content substantially lower than the current measurement, this
would provide evidence for an energy reservoir in the corona other
than thermal energy; a good candidate for such a reservoir would be
the magnetic field (Merloni \& Fabian 2001).

The total energy emitted during the hard state observations is about
$4\times 10^{40}$ ergs.  The long bursts seen in the island/hard
states of accreting neutron star systems are thought to be caused by
the nuclear burning of hydrogen (Muno et al. 2000).  Assuming an
accretion efficiency of about 20\%, about $2\times10^{20}$ grams of
material was accreted during that time.  Assuming solar composition,
about $1.5\times10^{20}$ grams must be in the form of hydrogen, which
would allow for about $10^{39}$ ergs to be produced by thermonuclear
burning of hydrogen into helium, assuming there is no steady burning.
Since the observed burst had a fluence of $3\times10^{38}$ ergs, the
expected number of bursts is about 3.  If only 10\% of the matter
actually reaches the surface of the neutron star, as claimed in
propeller model for the hard state (Zhang, Yu \& Zhang 1998), the
expected number of observed bursts would be about 0.3 (assuming bursts
can occur in polar accretion models, which is uncertain); thus, the
observation of a single burst does not place any tight constraints on
how much matter can be propelled away from the accretion flow.
Observations of bursts at lower luminosities might prove useful for
constraining the luminosity at which propeller effects set in.

\begin{figure*}
\centerline{\epsfig{figure=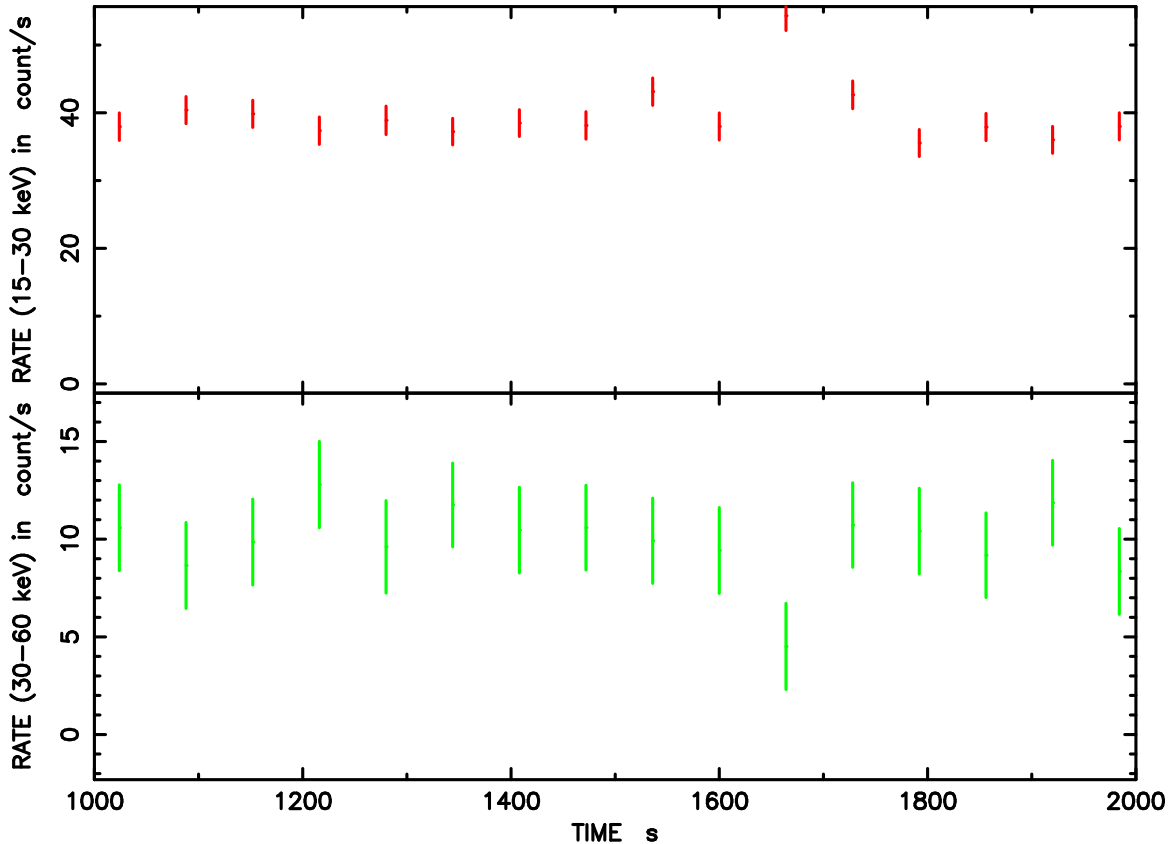,width=12 cm,angle=-90}}
\caption{The quenching of the corona during a type I X-ray burst in
Observation 40047-01-01-01.  The top panel shows the 15-30 keV HEXTE
lightcurve, while the bottom panel shows the 30-60 keV HEXTE
lightcurve.  At t=1650 seconds along the x-axis, the 15-30 keV
band shows its flux rising due to the burst, while the 30-60 keV flux
is quenched.}
\end{figure*}

\section{Conclusions}

We have found that the spectra of Aql X-1 in all spectral states can
be well fit by a pure thermal Comptonization model, but that other
models, including those with substantial non-thermal electron
components can also fit the data.  We have find a decrease in the
30-60 keV emission coincident with a type I burst in the hard state of
Aql X-1 and shown that such bursts can be used as probes of the
corona's total energy content and size.

\section{Acknowledgments}
We wish to thank Andrea Merloni, Mike Nowak and the anonymous referee
for useful comments.  This research has made use of data obtained from
the High Energy Astrophysics Science Archive Research Center
(HEASARC), provided by NASA's Goddard Space Flight Center.

\end{document}